\documentclass[draft,aps,twocolumn,showpacs,floats,prl,
superscriptaddress]{revtex4}  

\usepackage{psfig}
\newcommand{\stp}{{\tilde{t}_1}}
\newcommand{\mstp}{{m_{\tilde{t}_1}}}

\newcommand{\sbt}{{\tilde{b}_1}}
\newcommand{\msbt}{m_{\tilde{b}_1}}

\newcommand{\msbts}{m_{\tilde{b}_2}}
\newcommand{\glu}{{\tilde{g}}}
\newcommand{\mglu}{m_{\tilde{g}}}
\newcommand{\cha}{{\tilde{\chi}^-_1}}

\begin{document}

\preprint{KEK-TH-818}

\preprint{ICRR-Reprot-486-2002-4}

\preprint{KOBE HEP 02-01}

\preprint{YITP-02-05}
\title{Scenery from the Top: 
Study of the Third Generation Squarks at CERN LHC}
\author{Junji Hisano} 
\affiliation{ICRR, University of Tokyo,
Kashiwa 277-8582, Japan }
\affiliation{IPNS, KEK, Tsukuba 305-0801, Japan}
\author{Kiyotomo Kawagoe}
\affiliation{Department of Physics,
Kobe University, Kobe 657-8501, Japan}
\author{Ryuichiro Kitano}
\affiliation{IPNS, KEK, Tsukuba 305-0801, Japan}
\affiliation{Department of Physics, Tohoku University, Sendai 980-77, Japan}
\author{Mihoko M. Nojiri}
\affiliation{YITP, Kyoto University, Kyoto 606-8502, Japan}
\date{\today}

\begin{abstract}
In the minimal supersymmetric standard model (MSSM) properties of the
third generation sfermions are important from the viewpoint of
discriminating the SUSY breaking models and in the determination of
the Higgs boson mass. If gluinos are copiously produced at CERN LHC,
gluino decays into $tb$ through stop and sbottom can be studied using
hadronic decays of the top quark.  The kinematical endpoint of the
gluino decays can be evaluated using a $W$ sideband method to estimate
combinatorial backgrounds.  This implies that fundamental parameters
related to the third generation squarks can be reliably measured.  The
top-quark polarization dependence in the decay process may also be
extracted by looking at the $b$ jet distribution near the kinematical
endpoint.

\end{abstract}


\maketitle

The minimal supersymmetric standard model (MSSM) is one of the
promising extensions of the standard model (SM). The model requires
superpartners of ordinary particles (sparticles), and LHC at CERN
might confirm the existence of these new particles \cite{TDR}.  Since
the sparticle masses are related to the SUSY breaking mechanism,
measurement of the masses provides a way to probe the origin of SUSY
breaking in nature.

The masses and mixings of stops ($\tilde{t}_{1,2}$) and sbottoms
($\tilde{b}_{1,2}$) are sensitive to the flavor structure of scalar
masses. First, the diagonal masses in the stop and sbottom mass
matrices, $m_{\tilde{Q}_3}$, $m_{\tilde{t}_R}$, and $m_{\tilde{b}_R}$,
are predicted to be less than those of other squarks in the minimal
supergravity (MSUGRA) model \cite{SUGRA} because of the Yukawa RGE
running effects. Some SUSY breaking models, such as the flavor U(2)
model \cite{u2model} or the decoupling solution \cite{decoupling}, and
the superconformal model \cite{Nelson:2000sn} for the SUSY flavor problem,
have also special imprints on these parameters. The
$\tilde{t}_L$-$\tilde{t}_R$-Higgs trilinear coupling $A_t$ at the weak
scale has a large coefficient proportional to $m_t$, resulting in a
large left-right mixing of stops; $m_{LR}^2=
m_t(A_t-\mu\cot\beta)$. In the MSUGRA $A_t$ is proportional to the
universal gaugino mass $M$ at the GUT scale $M_{GUT}$, and insensitive
to the $A$ parameter at $M_{GUT}$~\cite{infrared}. Indeed, this is one
of the robust predictions of SUSY breaking at $M_{GUT}$. These
relation is not guaranteed if the SUSY breaking mediation scale is
much smaller than $M_{GUT}$. It should be stressed that the stop
masses and the mixing are very important parameters to predict the
light Higgs mass
\cite{higgsmass}.

It is possible to study the stop and sbottom at LHC through the gluino 
($\tilde{g}$) 
decay modes listed below.
\begin{eqnarray}
{\rm I)}&~&\glu\rightarrow b\sbt
\rightarrow bb\tilde{\chi}^0_2
\rightarrow bbl^+l^-\tilde{\chi}^0_1,\cr
{\rm II)}&~&\glu\rightarrow t\stp^*\rightarrow tt\tilde{\chi}^0_1,\cr
{\rm III)}&~&\glu\rightarrow t\stp^*\rightarrow tb\cha,\cr
{\rm IV)}&~&\glu\rightarrow \bar{b}\sbt\rightarrow tb \cha .
\label{gluinodecay}
\end{eqnarray}
In previous literatures~\cite{TDR,Hinchliffe:1998zj}, the third
generation sfermions are often studied using the mode I) ($bbl^+l^-$
channels). This mode is important when $\tilde{\chi}^0_2$ has
substantial branching ratios into leptons. Measurement of the
kinematical endpoints of the signal distributions tells us the
sparticle masses. Unfortunately, the branching ratios could be very
small, and this mode is insensitive to the stop. We tried to study the
mode II) in Eqs.~(\ref{gluinodecay}), but the result was unsuccessful
because of the small branching ratio in the MSUGRA.

In this paper we try to measure the endpoints of the modes III) and
IV) in Eqs.~(\ref{gluinodecay}). The decay modes are expected to be
dominant in the MSUGRA since $\cha$ is likely to be SU(2)$_L$
gaugino-like and $Br(\sbt,\stp\rightarrow\cha)$ could be as large as
60\%.  We focus on the reconstruction of hadronic decay of the top
quark, because the distribution of the $tb$ invariant mass $m_{tb}$
makes a clear endpoint in this case.

The parton level distribution of the $tb$ final state invariant mass
is expressed as a function of $\mglu$, $m_{\tilde{t}_1}$,
$m_{\tilde{b}_1}$, and chargino mass $m_{\cha}$: $d
\Gamma/dm_{tb}\propto m_{tb}$, and the endpoints $M_{tb}$ for the
modes III) and IV) are written as follows;
\begin{eqnarray}
&&M_{tb}^2({\rm III}) =  m_t^2
+\frac{m_\stp^2-m_\cha^2}{2 m_\stp^2}
\left\{
(m_\glu^2-m_\stp^2-m_t^2)
\right.
\nonumber\\
&&
\left.
+
\sqrt{
(m_\glu^2-(m_\stp-m_t)^2)
(m_\glu^2-(m_\stp+m_t)^2)
}
\right\},\nonumber
\label{tb_stop}
\\
&&M_{tb}^2({\rm IV})=  m_t^2
+\frac{m_\glu^2-m_\sbt^2}{2 m_\sbt^2}
\left\{
(m_\sbt^2-m_\cha^2+m_t^2)
\right.
\nonumber\\
&&
\left.
+
\sqrt{
(m_\sbt^2-(m_\cha-m_t)^2)
(m_\sbt^2-(m_\cha+m_t)^2)
}
\right\}. \nonumber
\label{tb_sbot}
\end{eqnarray}
Note that the endpoint of the final state $tb\cha$  should  be
sensitive to both $\tilde{t}_1$ and $\tilde{b}_1$.

In order to demonstrate the endpoint reconstruction, we take an MSUGRA
point with the scalar mass $m=100$GeV, the gaugino mass $M=300$GeV,
the $A$ parameter $-300$GeV at the GUT scale, $\tan\beta=10$ and
$\mu>0$. This corresponds to the sample point A1 in Table~1. The
masses and mixings of sparticles are calculated by ISASUSY/ISASUGRA
\cite{ISAJET}. The Monte Carlo SUSY events are generated by PYTHIA
\cite{PYTHIA} using the masses and mixings, and then passed through a
fast detector simulation program for the ATLAS experiment
\cite{ATLFAST}.  Jets are reconstructed by a cone-based algorithm with
$\Delta R=0.4$.  The $b$ and $\tau$ tagging efficiencies are set to be
60\% and 50\%, respectively.

\begin{table}[t]
\begin{tabular}{|c|c|c|c|c|c||c|c|c|c|c|c|
}
\hline
 & $\mglu$&$\mstp$&$\msbt$& $\msbts$ &$\sigma$&
 & $\mglu$&$\mstp$&$\msbt$& $\msbts$ &$\sigma$\cr
\hline
A1 & 707& 427 & 570 & 613 &26 &
A2 & 706& 496 & 587 & 614&25\cr
G1& 807& 427 & 570 & 613&18 &
G2 & 806& 496& 587& 614 & 18\cr
T1&707&327&570&613&30 &
T2 &707&477&570&612&25\cr
\hline
B & 609& 402 & 504 & 534 & 56&
C & 931& 636 & 771 & 805 & 5\cr
G & 886& 604 & 714 & 763 & 7 &
I & 831& 571& 648 & 725 & 10\cr
\hline
E1      & 515& 273& 521& 634& 77 &
E2      & 747& 524 & 770& 898& 8\cr
\hline
\end{tabular}
\caption{The sparticle masses in GeV and the total SUSY cross
sections in pb for the parameter 
points studied in this paper. }
\end{table}

In our study we apply the following selection for the $tb$ signal; 1)
$E_T^{\rm miss} > 200$~GeV, 2) $m_{\rm eff} > 1000$~GeV ($m_{\rm eff}
= E_T^{\rm miss} + \sum_{all}{p_T^{\rm jet}} $), 3) two and only two
$b$-jets with $p_T > 30$~GeV, 4) $4 \leq n_{\rm jet} \leq 6$ ($n_{\rm
jet},$ number of additional jets with $p_T > 30$~GeV and
$|\eta|<3.0$).  In addition, events with leptons are removed to reduce
the background from $t\bar{t}$ production. At this stage the number of
remaining $t\bar{t}$ events is rather small, about 10\% of the
remaining SUSY events for the point A1.

To reconstruct the hadronic decay of the top quark, we first take a
jet pair consistent with a hadronic $W$ boson decay with a cut on the
jet pair invariant mass $m_{jj}$; $\vert m_{jj}-m_W\vert<15$GeV. The
invariant mass of the jet pair and one of the $b$ jets, $m_{bjj}$, is
then calculated.  All possible combinations are tried in an event, and
the combination which minimizes $| m_{bjj}-m_t |$ is chosen.  The jet
combination is regarded as a top candidate if $\vert m_{bjj}-m_t\vert<
30$GeV.  The energy and momentum of the jet pair are then rescaled so
that $m_{jj}=m_W$.

The expected $tb$ endpoint is not clearly visible in the $m_{tb}$
distribution shown in Fig.~1(a). As there are 7 to 8 jets on average
in a selected event, events with a fake $W$ boson (and a fake top
quark) dominate the distribution. The distribution of the fake $W$
events is estimated from the events that contain jet pairs with the
invariant mass in the ranges A) $\vert m_{jj}-(m_W-15{\rm
GeV})\vert<15$ GeV and B) $\vert m_{jj}-(m_W+15{\rm GeV})\vert<15$
GeV; `the $W$ sidebands'.  The energy and momentum of the jet pairs
are then rescaled linearly to be in the range $\vert
m_{jj}-m_W\vert<15$GeV. The fake top candidates are reconstructed from
the rescaled jet pairs and $b$ jets in the events.

\begin{figure}[t]
\centerline{\psfig{file=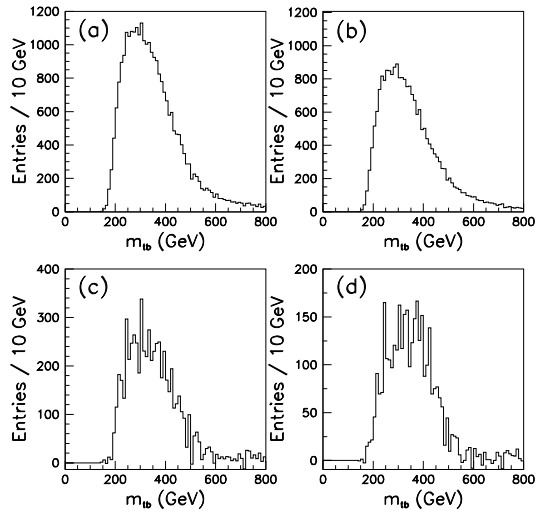,width=3in}}
\caption{
(a) The signal $m_{tb}$ distribution for the sample point A1 in
Table~1, (b) the estimated background distribution from the sideband
events,  (c) (a)$-$(b), and (d) the $m_{tb}$ distribution for the modes III)
and IV) in Eqs.~(\ref{gluinodecay}), and a decay mode 
irreducible to the mode III).
}
\label{fig1}
\end{figure}

The estimated background distribution is shown in Fig.~1(b), which is
obtained by averaging distributions from the sidebands A) and B).  The
estimation is based on an assumption that most of the jets in the
events do not have significant correlation with the $b$ jets in the
events.  The estimated background distribution is subtracted from the
signal distribution in Fig.~1(c).  The estimated {\it correct} signal
distribution (c) shows the better endpoint compared to (a). Fig.~1(d)
is the same distribution as Fig.~1(c) but for the events which contain
the mode III), the mode IV), or a decay mode irreducible to the mode
III); $\glu\rightarrow \sbt b \rightarrow \stp (W b)\rightarrow b\cha
(bW)$. Note that if $(bW)$ has an invariant mass consistent to a top,
the decay is kinematically equivalent to the mode III).  Fig.~1(d)
shows the expected clear edge at the right place
($M_{tb}({\mathrm{III}})=476$GeV and $M_{tb}({\mathrm{IV}})=420$ GeV),
demonstrating that the sideband method works well. Here, the number of
the generated SUSY events is $3\times10^6$, which corresponds to an
integrated luminosity of $120$fb$^{-1}$. The plots do not include the
SM backgrounds.

Note that the signals from the modes III) and IV) in
Eqs.~(\ref{gluinodecay}) are significant in the total selected
events. We fit the total distribution shown in Fig.~1(c) by a simple
fitting function, which is described as a function of the endpoint
$M_{tb}^{\rm fit}$, the edge height $h$, and the smearing parameter
$\sigma$ from the jet energy resolution. We assume that the signal
distribution is sitting on a linearly-decreasing background. The
$M^{\rm fit}_{tb}$ is compared with the weighted endpoint
$M_{tb}^{\rm w}$ defined by
$$
M_{tb}^{\rm w}=\frac{Br({\rm III})  M_{tb}({\rm III}) 
+Br({\rm IV}) M_{tb}({\rm IV})}
{Br({\rm III})+ Br({\rm IV})}, 
$$
where $Br({\rm III})$ and $Br({\rm IV})$ are branching ratios for the
modes III) and IV), respectively.  The fit is shown in Fig.~2 (a). We
obtain $M^{\rm fit}_{tb}=443.2\pm 7.4$ GeV, which is consistent with $
M^{\rm w}_{tb}=459$GeV. The $M^{\rm fit}_{tb}$ changes moderately when
one changes the $m_{tb}$ range used for the fit. We choose the range
so that the significance of the height $S=h/\Delta h$ is at maximum.
For the fit shown in Fig.~2(a), $S({\rm max})=196.9/15.2=13.0$ is
obtained.

\begin{figure}[t]
\centerline{\psfig{file=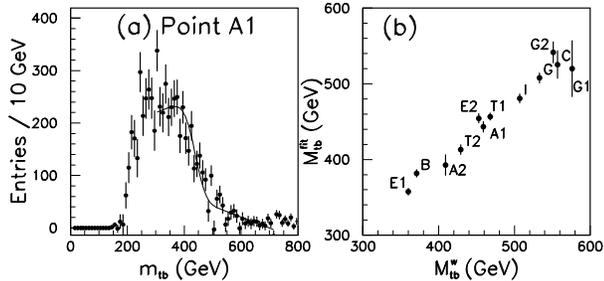,width=3.4in}}
\caption{(a) A fit to the $m_{tb}$ distribution (point A1), and
(b) comparison of $M^{\rm fit}_{tb}$ and $M^{\rm w}_{tb}$ for the sample
parameters given in this paper. }
\label{fig2}
\end{figure}

In order to check the availability of the endpoint measurement, we
study twelve sample points in total, including the previous point,
shown in Table~1, and compare $M^{\rm fit}_{tb}$ and $M^{\rm w}_{tb}$. 
We choose two reference MSUGRA points as A1 and A2, where $m=100$ GeV,
$M= 300$ GeV, $\tan\beta=10$, $\mu>0$ and $A=\mp 300$ GeV. We also
study points with different mass spectrums from the MSUGRA
predictions; two points with gluino masses heavier than the reference
points, (G1, G2), and two points with modified stop masses
(T1,T2). Furthermore, we include the MSUGRA points selected from
\cite{ellis} (B,C,G,I) and two non-SUGRA points E1 and E2 where the
gluino decays exclusively into $\tilde{t}_1$. The result is summarized
in Fig.~2 (b)\footnote{ For the point C where SUSY decays contain many
leptons, we omit the lepton veto and reduce the $t\bar{t}$ background
by requiring $m_{bl}>140$ for any lepton with $p_T>10$~GeV.  }. The
error bars represent the statistical errors for $3 \times 10^6$ SUSY
events at each point, and the systematic error of the jet energy scale
(1\%) is not included (see Table~1 for SUSY cross sections). This plot
shows an impressive linearity between the expectation and the MC fits,
although $M^{\rm fit}_{tb}$ is systematically lower than $M^{\rm
w}_{tb}$. This is reasonable since some of particles are always out of
the cone to define jets. This effect may be corrected by comparing
distributions with different jet definitions.

Another uncertainty may come from the jet fragmentation. If events are
generated by ISAJET for the point A1, the reconstructed endpoint is
smaller by 10\%, and the number of events after the sideband
subtraction is smaller by a factor of 1.5. The difference comes from
the different jet fragmentation schemes. ISAJET radiates more soft
jets for a parton, resulting in more background and smeared endpoint
distribution. The event generators must be tuned carefully to extract
the kinematical information from the signal distribution.

We now discuss the physics that might be studied with the $tb$ endpoint
measurement.

We cannot determine all of the relevant mass parameters from only the
$tb$ endpoint measurement, since the endpoint $M_{tb}$ depends on
$\mglu$, $\msbt$, and $\mstp$. Thus, the study of $bbl^+l^-$ final
state would be important to single out the possible $\tilde{t}$ and
$\tilde{b}$ contributions to the $tb$ final state and to proceed to a
model-independent study.  For example, let us assume that errors of
the endpoints $M_{bbll}$ and $M_{bll}$ are 10 GeV and that of the
endpoint $M_{b_1l}$ is 30 GeV in the measurement. Here, $b_1$ is one
of the two $b$ jets for which the invariant mass $m_{bll}$ is
larger. When we generate $\mglu$, $\msbt$, $\tilde{\chi}^0_2$, and
$\tilde{\chi}^0_1$ randomly around the reference point A1 fixing the
endpoint $M_{ll}$ (which is expected to have a very small error), and
require $\Delta\chi^2\equiv$ $\sum_i (M_{i}-M^{A1}_{i})^2/\Delta
M_{i}^2$ $<1(9)$ ($i$ runs over the possible endpoints), the deviation
$M_{tb} ({\rm IV}) - M^{A1}_{tb} ({\rm IV})$ is always less than
15(45)GeV. For the point A1, $M_{tb}({\rm IV})=420$ GeV and $M^{\rm
w}_{tb}\sim 460$ GeV. The difference of $M_{tb}({\rm IV})$ and $M^{\rm
w}_{tb}$ therefore may be statistically distinguishable.

The measurement of the SUSY breaking parameters in different sectors
might reveal an overall inconsistency of the SUSY breaking mediation
models.  The distribution of the invariant mass formed by combining
the highest $P_T$ jet and a same-flavor and opposite-sign lepton pair
($jll$ channels) is sensitive to $m_{\tilde{q}}$ and $m_{\tilde{l}}$,
and this may lead to the determination of $m$ and $M$ in the MSUGRA
\cite{TDR,jll}. Once $M$ and $m$ are fixed, they strongly constrain
the endpoint $M^{\rm w}_{tb}$ in the MSUGRA -- by comparing it with
the measured value, one should be able to check if $m_{\tilde{b}}$ or
$m_{\tilde{t}}$ are consistent to the MSUGRA predictions.

Note also that the formulae of $M_{tb}$ or $M_{bbll}$ involve
$m_{\tilde{g}}$, while the first generation squark mass is lower
bounded by $m_{\tilde{g}}$ in the model. If $\mglu\gg m_{\tilde{q}}$
is established by combining the squark mass scale determined through a
$jll$ analysis~\cite{TDR} and the analysis of the final state
involving $b$ jets, we can show that some new physics should occur
below the scale where $m^2_{\tilde{q}}<0$.  Note that, in the points
G1 and G2 in Table 1, where the gluino mass is increased by 100 GeV
from the MSUGRA predictions, $m^2$ becomes negative at the GUT
scale. We study the $jll$ distributions for the point G2 in a similar
way as given in Ref.~\cite{jll} and find that the $jll$ endpoints are
successfully reconstructed.  Therefore the information on
$m_{\tilde{q}}$ should be obtained in this case.

\begin{figure}[t]
\centerline{\psfig{file=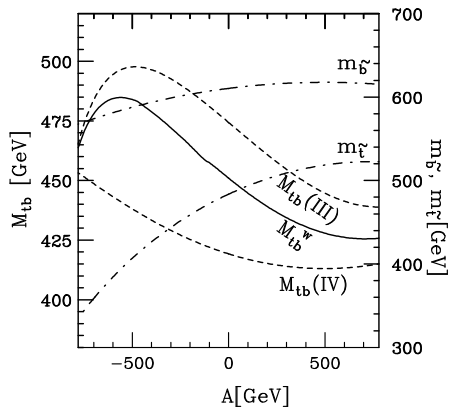,width=2.3in}}
\caption{Dependence of $M_{tb}$, $m_{\tilde{t}_1}$, and $m_{\tilde{b}_1}$
on the GUT scale A parameter in the MSUGRA.} 
\label{fig3}
\end{figure}

In the framework of the MSUGRA, the measurement of $M_{tb}$ is
sensitive to the GUT scale $A$ parameter. This effect is large when
$M\times A<0$ as can be seen in Fig.~3.  Here we take $M\sim m$ and
$m=230$ GeV, and vary $A$ so that $\vert A\vert< 3m$.  The $M^{\rm
w}_{tb}$ and $\mstp$ vary by 50 GeV and 150 GeV, respectively, and the
changes are again detectable. Note that the $\tilde{\chi}^0_2$ decay
into $\tilde{l}l$ is closed in this case, therefore the information
from the $bbl^+l^-$ channels is not available. If $m=100$ GeV and $A$
is varied from $-300$ GeV to 300 GeV, $M^{\rm w}_{tb}$ and $\mstp$
change by 30 GeV and 70 GeV, respectively.

\begin{figure}[bt]
\centerline{\psfig{file=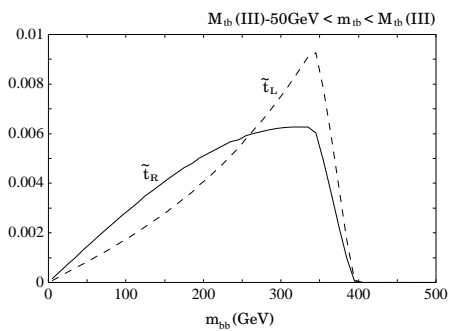,width=2.3in}}
\caption{The
parton level distribution of $m_{bb}$ for point A1. The two histograms
show the distribution for the event with $M_{tb}({\rm III})-50{\rm
GeV}<m_{tb}<M_{tb}({\rm III})$ in cases $\tilde{t}_1=\tilde{t}_R$ and
$\tilde{t}_1=\tilde{t}_L$.
}
\label{fig4}
\end{figure}

Finally, we discuss top-quark polarization dependence in the gluino
decay.  Naturally all top quarks coming from $\glu$, $\stp$, and
$\sbt$ decays are polarized, depending on the sfermion and chargino
mixings.  For example, when $\tilde{t}_1$ is mostly right(left)-handed
stop, $t$ of $\glu\rightarrow t\tilde{t}_1^*$ is $t_{R(L)}$. This
propagates into the average top-quark helicity, if the top quark is
relativistic enough in the gluino rest frame.  If the $tb$ invariant
mass is close to the endpoint $M_{tb}$ of the mode III) in
Eqs.~(\ref{gluinodecay}), the $t$ and $b$ go to the opposite direction
in the gluino rest frame. Because the bottom quark from the top-quark
decay tends to go to the opposite direction to the top-quark helicity,
the distribution of the invariant mass $m_{bb}$ for events with
$m_{tb}$ close to $M_{tb}({\rm III})$ depends on the average top-quark
helicity. The parton-level distributions of $m_{bb}$ are plotted in
Fig.~4 for the point A1. The solid (dotted) curve shows the
distribution for the events with $M_{tb}({\rm III})-50{\rm
GeV}<m_{tb}<M_{tb}({\rm III})$ in the case of
$\tilde{t}_1=\tilde{t}_R$ ($\tilde{t}_1=\tilde{t}_L$). One can read
from Fig.~1 that roughly 200 signal events are available near the
endpoint.  The statistical difference between the two distributions is
3 $\sigma$ for 100 signal events without taking care of the background
\footnote{
For events with
$m_{tb} < M_{tb}(III)-50$ GeV the gluino decay 
into $\sbt$ starts to contribute.
}.

In this paper, we try to reconstruct final states which consisting of
hadronic jets at LHC. This was considered difficult due to the large
combinatorial background, but is overcome by a $W$ sideband method
developed to estimate the background.  We reconstruct the $tb$ final
state from the event containing two $b$ jets. The reconstructed
endpoint provides us an access to the gluino and the third generation
sparticle masses without relying on the decay modes including leptons.
The correct reconstruction of the events also allows us to consider
the top-quark polarization dependence of the distribution. This
information is important to determine the radiative correction to the
Higgs mass, as well as the origin of SUSY breaking.

It is a pleasure to thank Dr. Kanzaki and Mr. Toya for the
construction of a simulation environment. We acknowledge ICEPP,
Univ. of Tokyo, for providing us computing resources. This work is
supported in part by the Grant-in-Aid for Science Research, Ministry
of Education, Science and Culture, Japan (No. 12047127 for MMN).


\begin{thebibliography}{99}

\bibitem{TDR}
	ATLAS: Detector and physics performance technical design report,
	CERN-LHCC-99-14.


\bibitem{SUGRA}
See references in H.~P.~Nilles,
Phys.\ Rept.\  {\bf 110}, 1 (1984).

\bibitem{u2model}
R.~Barbieri, G.~R.~Dvali and L.~J.~Hall,
Phys.\ Lett.\ B {\bf 377}, 76 (1996).

\bibitem{decoupling}
See references in J.~Hisano, K.~Kurosawa and Y.~Nomura, 
Phys.\ Lett.\ B {\bf 445}, 316 (1999).

\bibitem{Nelson:2000sn}
A.~E.~Nelson and M.~J.~Strassler,
JHEP {\bf 0009}, 030 (2000);
T.~Kobayashi, H.~Nakano, T.~Noguchi and H.~Terao,
hep-ph/0202023.

\bibitem{infrared}
M.~Carena and C.~E.~Wagner,
Nucl.\ Phys.\ B {\bf 452}, 45 (1995).

\bibitem{higgsmass}
H.~E.~Haber and R.~Hempfling,
Phys.\ Rev.\ Lett.\  {\bf 66}, 1815 (1991);
Y.~Okada, M.~Yamaguchi and T.~Yanagida,
Prog.\ Theor.\ Phys.\  {\bf 85}, 1 (1991);
J.~R.~Ellis, G.~Ridolfi and F.~Zwirner,
Phys.\ Lett.\ B {\bf 257}, 83 (1991);
R.~Barbieri and M.~Frigeni,
Phys.\ Lett.\ B {\bf 258}, 395 (1991).

\bibitem{Hinchliffe:1998zj}
I.~Hinchliffe, F.~E.~Paige, E.~Nagy, M.~D.~Shapiro, J.~Soderqvist and W.~Yao,
LBNL-40954.

\bibitem{ISAJET}
H. Baer, F. E. Paige, S. D. Protopopescu and  X. Tata 
``ISAJET 7.48: a Monte Carlo event generator for $pp$, 
$\bar{p}p$, and $e^+ e^-$ reactions'',
hep-ph/0001086.

\bibitem{PYTHIA}
T. Sjostrand, L. Lonnblad and S. Mrenna 
``PYTHIA 6.2: physics and manual'',
hep-ph/0108264.

\bibitem{ATLFAST}
E.Richter-Was et al., ATLFAST2.21, ATLAS Internal Note PHYS-NO-079.

\bibitem{ellis}
M.~Battaglia {\it et al.},
Eur.\ Phys.\ J.\ C {\bf 22}, 535 (2001).

\bibitem{jll}
H.~Bachacou, I.~Hinchliffe and F.~E.~Paige,
Phys.\ Rev.\ D {\bf 62}, 015009 (2000).

\end{thebibliography}
\end{document}